\documentclass{PoS}

\usepackage{amsmath}

\newcommand{\be}{\begin{eqnarray}}
\newcommand{\ee}{\end{eqnarray}}
\newcommand{\lpar}{\left(}
\newcommand{\rpar}{\right)}
\newcommand{\re}{{\rm Re}}
\newcommand{\tr}{{\rm Tr}}
\newcommand{\conj}{\dagger}
\newcommand{\fermion}{\chi}
\newcommand{\q}{\fermion}
\newcommand{\qbar}{\bar\fermion}

\newcommand{\Eq}[1]{Eq.~\eqref{#1}}
\newcommand{\Fig}[1]{Fig.~\ref{#1}}

\usepackage{textpos}
\newcommand{\preprint}[1]{
	\begin{textblock}{1}(0,-3.4)
		\texttt{\hfill\footnotesize #1}
	\end{textblock}
}

\title{Integrating out lattice gauge fields 
\preprint{CERN-PH-TH-2015-140}
}

\ShortTitle{Integrating out lattice gauge fields}

\author{\speaker{H\'{e}lvio Vairinhos}\\
        Institute for Theoretical Physics, ETH Z\"{u}rich, CH-8093 Z\"{u}rich, Switzerland\\
        E-mail: \email{helviov@phys.ethz.ch}}

\author{Philippe de Forcrand\\
        Institute for Theoretical Physics, ETH Z\"{u}rich, CH-8093 Z\"{u}rich, Switzerland\\
        CERN, Physics Department, TH Unit, CH-1211 Geneva 23, Switzerland\\
        E-mail: \email{forcrand@phys.ethz.ch}}

\abstract{
The sign problem is a major obstacle to our understanding of the phase diagram of QCD at finite baryon density. Several numerical methods have been proposed to tackle this problem, but a full solution to the sign problem is still elusive. Motivated by this problem and by recent advances in diagrammatic Monte Carlo methods, we find a new exact representation of the partition function of pure lattice gauge theory that contains no link variables. This approach can be easily extended to include staggered fermions, and results in a diagrammatic representation of fermionic states as arrangements of monomers, dimers, and fermionic loops saturating the spacetime lattice. Our representations are exact for any value of the lattice coupling, and extend previous representations that are only valid in the strong coupling limit and at $O(\beta)$. As a concrete example, we construct a monomer-dimer-loop representation of compact lattice QED.
}

\FullConference{9th International Workshop on Critical Point and Onset of Deconfinement - CPOD2014,\\
		17-21 November 2014\\
		ZiF (Center of Interdisciplinary Research), University of Bielefeld, Germany}

\begin{document}

\section{Introduction}

The quantitative understanding of the phase diagram of nuclear matter is a very important open problem which defies both analytical and numerical treatments \cite{forcrand}. On the analytical side, sketches of the QCD phase diagram can be drawn from mean field arguments and the analysis of simple truncated models, which are mostly conjectural. On the numerical side, the notorious sign problem prevents the use of standard Monte Carlo methods to simulate lattice QCD at finite baryon density from first principles. 

Traditionally, the fermion fields in the partition function of lattice QCD are integrated  analytically a priori, which reduces the system to an ensemble of color-charged gauge fields weighted by the determinant of the lattice Dirac operator. The sign problem occurs for finite values of the baryon chemical potential, for which such a determinant is, in general, complex-valued. The direct Monte Carlo sampling of this ensemble is therefore impossible. Several numerical methods have been developed to deal with this difficulty, but none has succeeded yet in providing a satisfactory description of the QCD phase diagram.

The common denominator to all such methods is the a priori integration over the lattice fermions, and the subsequent sampling over degrees of freedom carrying color-charge. However, QCD is known to be confining, in the sense that its asymptotic eigenstates are color-neutral. Hence, it would be more ``natural'' for the Monte Carlo integration to sample over color-neutral degrees of freedom -- at least in the confining regime -- rather than sampling over local gauge-covariant states, whose quantum fluctuations and local gauge ambiguity propagate into uncontrollably large fluctuations of the phase of the Dirac determinant. More precisely, if $|\psi_i\rangle$ denotes the putative complete basis of eigenstates of the Euclidean QCD Hamiltonian $H$,
\be
Z = \tr(e^{-\beta H}) = \tr\left( e^{-\frac{\beta H}{N_t}}
	\sum_i |\psi_i\rangle\langle\psi_i| e^{-\frac{\beta H}{N_t}}
	\sum_j |\psi_j\rangle\langle\psi_j| e^{-\frac{\beta H}{N_t}} \cdots\right)
\ee
then all the matrix elements $\langle\psi_i| e^{-\frac{\beta H}{N_t}} |\psi_j\rangle$ are non-negative by definition, hence the sign problem disappears in such a basis. In short, the severity, and even the existence, of a sign problem depends on the particular representation of the partition function, $Z$.

Even though it is unfeasible to represent the partition function of lattice QCD in terms of its physical eigenstates, it is reasonable to think that the sign problem might be significantly reduced when $Z$ is represented in terms of color-neutral states which are not necessarily eigenstates. Such a representation can be obtained by integrating the lattice gauge fields a priori. Let us consider the partition function of $SU(N)$ or $U(N)$ lattice gauge theory with a single flavor of staggered fermions,
\be
Z = \int [dU][d\qbar d\q]\, 
	e^{-\beta\sum_p\lpar 1-\frac{1}{N} \re\tr(U_p) \rpar} 
	e^{\sum_{x,\mu} \eta_{x\mu} \tr(\qbar_x U_{x\mu} \q_{x+\hat\mu}
	- \qbar_{x+\hat\mu} U_{x\mu}^\dag \q_x)}
	e^{2am \sum_{x} \qbar_x\q_x} 
\ee
where $U_{x\mu}$ are link variables, $U_p$ is the path-ordered product of link variables around a plaquette, $\qbar_x,\q_x$ are staggered fermions, $am$ is their bare mass in lattice units, $\beta$ is the bare lattice coupling, and $\eta_{x\mu} = \pm 1$ are staggered phases. The direct analytic integration of the gauge fields is only feasible at $\beta=0$: the plaquette terms vanish, and the fermionic weight factorizes into a product of solvable one-link integrals \cite{mdp},
\be
Z &=& \int [d\qbar d\q]\, 
	e^{2am \sum_{x} \qbar_x\q_x} 
	\prod_{x,\mu} \int dU_{x\mu}\,
	e^{\eta_{x\mu} \tr(\qbar_x U_{x\mu} \q_{x+\hat\mu}
	- \qbar_{x+\hat\mu} U_{x\mu}^\dag \q_x)}
\\&=& \int [d\qbar d\q]\, 
	e^{2am \sum_{x} M_x} 
	\prod_{x,\mu} 
	\lpar
	\sum_{k=0}^N \frac{(N-k)!}{N!\,k!} M_x M_{x+\hat\mu} 
	+ \bar B_x B_{x+\hat\mu}
	+ (-1)^N \bar B_{x+\hat\mu} B_x
	\rpar
	\label{eq::Z::b0}
\ee
The resulting partition function is a Grassmann integral whose integrand contains color-neutral terms only, namely mesons $M_x = \qbar_x\q_x$ and baryons $B_x = \frac{1}{N!} \varepsilon_{i_1 \cdots i_N} \chi_x^{i_1} \cdots \chi_x^{i_N}$. Color-neutral states made of such building blocks are not exact eigenstates of lattice QCD, but they resemble the expected asymptotic states of the continuum theory.

\section{Dimer representation of lattice QCD}

\begin{figure}
\centering
\includegraphics[width=0.30\textwidth]{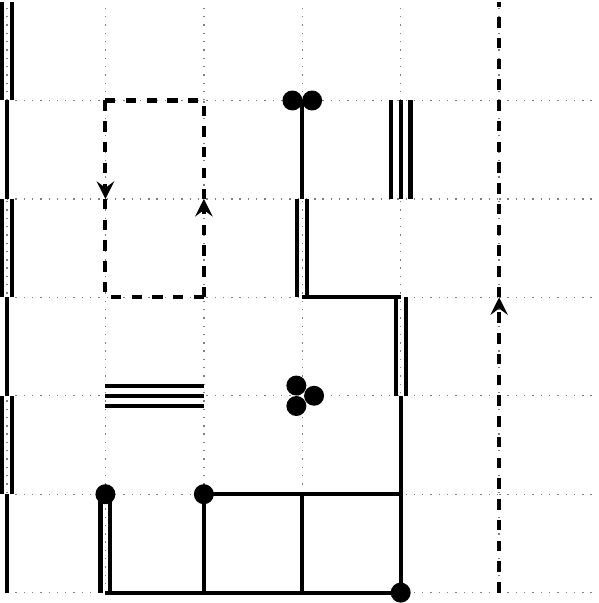}
\caption{Typical MDP configuration of $N_f=1$ lattice QCD on a $6 \times 6$ lattice, at $\beta=0$. \label{fig::mdp_su3}}
\end{figure}

The lattice fermions in \Eq{eq::Z::b0} still need to be integrated out. Solving the Grassmann integrals results in a combinatorial partition function \cite{mdp},
\be
Z = \sum_{\{n,k,\ell\}} 
	\prod_x \frac{N!}{n_x!} (2am)^{n_x}
	\prod_{x,\mu} \frac{(N-k_{x\mu})!}{N!\,k_{x\mu}!}\,
	\frac{\sigma(\ell)}{N!^{|\ell|}}
	\label{eq::Z::MDP}
\ee
whose degrees of freedom are monomers, dimers and self-avoiding polygons which saturate the spacetime lattice. More precisely:
\begin{itemize}
	\item $n_x \in \{0,\ldots,N\}$ represents the number of monomers occupying the lattice site $x$. It is interpreted as the condensate contribution $(M_x)^{n_x}$ to the fermionic state. The weight of such monomers depends on the bare quark mass $am$ and is positive-definite.
	\item $k_{x\mu} \in \{0,\ldots,N\}$ represents the number of dimers occupying the lattice bond $(x,\hat\mu)$. It is interpreted as the meson hopping contribution $(M_x M_{x+\hat\mu})^{k_{x\mu}}$ to the fermionic state. The weight of such dimers is also positive-definite.
	\item $\ell$ is an oriented self-avoiding polygon. It is interpreted as the contribution to the fermionic state of a sequence of baryon hoppings $\bar B_x B_{x+\hat\mu}$ which form a closed loop. The weight of a baryonic loop $\ell$ depends on its perimeter $|\ell|$ and on its sign $\sigma(\ell) = \pm 1$, which is a specific function of the topology and shape of $\ell$. The sign problem occurs due to the existence of baryonic loops $\ell$ for which $\sigma(\ell)=-1$.
\end{itemize}

The Grassmann integration constrains the monomer-dimer-polygon (MDP) configurations, only allowing those which saturate each site with exactly $N$ quarks and $N$ anti-quarks. Each baryonic loop must therefore be a self-avoiding polygon which saturates every site it crosses, while the number of monomers or dimers attached to the remaining sites must be exactly $N = n_x + \sum_{\pm\mu} k_{x\mu}, \; \forall x$ (see \Fig{fig::mdp_su3} for example).

Constrained combinatorial systems such as \Eq{eq::Z::MDP} can be simulated efficiently with variants of the worm algorithm \cite{worm}. Recent simulations of the MDP representation of lattice QCD ($N=3$) using such algorithms \cite{forcrand-mdp} show that the sign problem becomes very mild, with an improvement by a factor of $O(10^4)$ in accessible volumes over the traditional approach. The sign problem is also shown to remain mild when $O(\beta)$ corrections are included. Consequently, the phase diagram of lattice QCD in the strong coupling limit and its $O(\beta)$ corrections have been mapped successfully \cite{forcrand-mdp}.

\section{Decoupling the plaquette}

Higher-order corrections to the strong coupling limit of lattice QCD in the MDP representation involve combinatorially large numbers of cumbersome group integrals. In order to approach the regime of continuum physics, it would be desirable to use a MDP representation of lattice QCD which is not of the form of a strong coupling expansion. However, it is not possible to directly integrate out the link variables in the presence of Wilson action terms, which couple the four link variables around each plaquette. We circumvent this analytical difficulty by coupling the link variables to auxiliary fields in such a way that that links decouple among themselves \cite{nlink}. We illustrate this in the case of the pure gauge theory,
\be
Z = \int [dU]\, e^{-\beta\sum_p\lpar 1-\frac{1}{N} \re\tr(U_p) \rpar} 
\label{eq::Z::4link}
\ee
We refer to \Eq{eq::Z::4link} as the 4-link partition function, given that it only contains terms with four link variables coupled around a plaquette.

\begin{figure}
\centering
\includegraphics[width=0.18\textwidth]{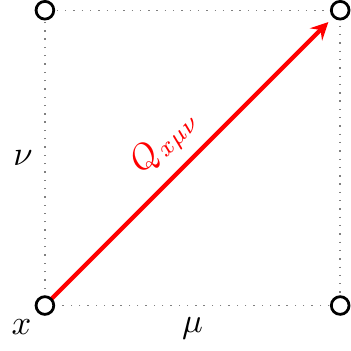} \qquad
\includegraphics[width=0.18\textwidth]{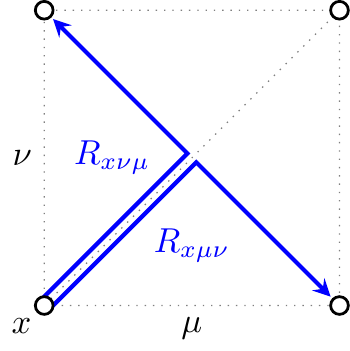}
\caption{Graphical representation of the $Q$ variables (left) and of the $R$ variables (right).\label{fig::QR}}
\end{figure}

\begin{figure}
\centering
\includegraphics[width=0.60\textwidth]{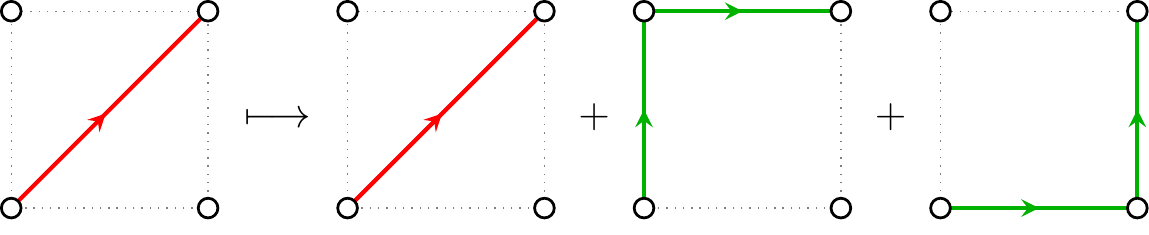} \vskip 5mm
\includegraphics[width=0.445\textwidth]{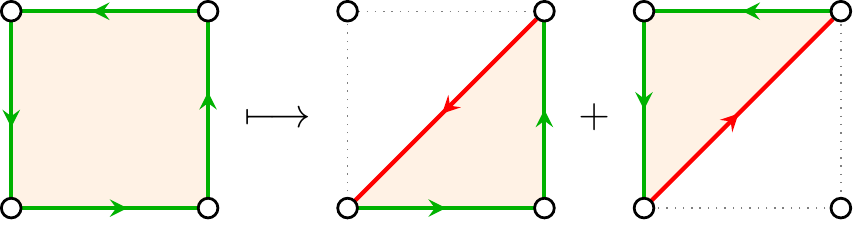}
\caption{Graphical representation of the HS transformation of the $Q$ variables (top), which induces the replacement of the Wilson plaquette terms in the 4-link partition function by $\frac{1}{2}$-plaquette terms (bottom). \label{fig::HS::Q}}
\end{figure}

First, we introduce a set of free auxiliary variables $Q_{x\mu\nu} \in \mathbb{C}^{N\times N}$ living on ``diagonal'' links (see \Fig{fig::QR}), i.e. with Gaussian distribution $dQ_{x\mu\nu}^\dag dQ_{x\mu\nu} \exp(-\frac{1}{2N} \tr(Q_{x\mu\nu}Q^\dag_{x\mu\nu}))$. A Hubbard-Stratonovich (HS) transformation of the form \cite{nlink,HS}:
\be
Q_{x\mu\nu} \mapsto \sqrt{\frac{\beta}{N}} 
(Q_{x\mu\nu} + U_{x\mu}U_{x+\hat\mu,\nu} + U_{x\nu}U_{x+\hat\nu,\mu})
\label{eq::HS::Q}
\ee
splits each Wilson plaquette term into two $\frac{1}{2}$-plaquette terms, each of which couples two link variables $U$ and one auxiliary variable $Q$ (see \Fig{fig::HS::Q}). The partition function \Eq{eq::Z::4link} then becomes:
\be
Z \propto \int [dQdQ^\dag]\, e^{-\frac{\beta}{2N} \tr(QQ^\dag)} 
\int [dU]\, e^{-\beta\sum_{x,\mu\neq\nu}\lpar 1-\frac{1}{N} 
\re\tr(Q_{x\mu\nu} U_{x\mu} U_{x+\hat\mu,\nu} ) \rpar} 
\label{eq::Z::2link}
\ee
which we refer to as the 2-link partition function, given that it only contains terms with two link variables coupled around a $\frac{1}{2}$-plaquette.

\begin{figure}
\centering
\includegraphics[width=0.60\textwidth]{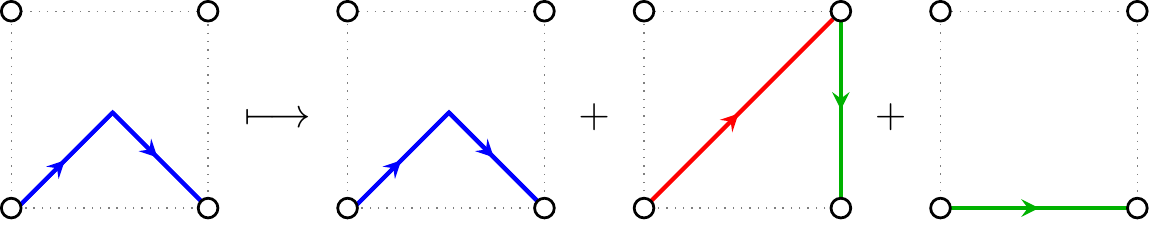} \vskip 5mm
\includegraphics[width=0.91\textwidth]{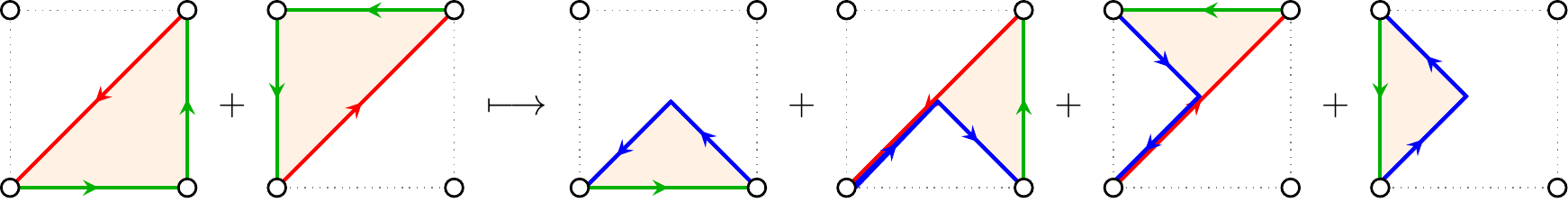}
\caption{Graphical representation of the HS transformation of the $R$ variables (top), which induces the replacement of the $\frac{1}{2}$-plaquette terms in the 2-link partition function by $\frac{1}{4}$-plaquette terms (bottom). \label{fig::HS::R}}
\end{figure}

Second, we introduce yet another set of free auxiliary variables $R_{x\mu\nu} \in \mathbb{C}^{N\times N}$ (i.e. with Gaussian distribution) living on ``folded'' links (see \Fig{fig::QR}). A HS transformation of the form \cite{nlink}:
\be
R_{x\mu\nu} \mapsto \sqrt{\frac{\beta}{N}} 
(R_{x\mu\nu} + Q_{x\mu\nu}U_{x+\hat\mu,\nu}^\dag + U_{x\mu})
\label{eq::HS::R}
\ee
splits each $\frac{1}{2}$-plaquette term of \Eq{eq::Z::2link} into two $\frac{1}{4}$-plaquette terms, each of which couples a single link variable $U$ to ambient auxiliary variables $Q$ and $R$ (see \Fig{fig::HS::R}). The partition function \Eq{eq::Z::2link} then becomes:
\be
Z \propto \int [dRdR^\dag][dQdQ^\dag]\, 
e^{-\frac{\beta}{2N}\tr(RR^\dag)} 
e^{-\frac{3\beta}{2N}\tr(QQ^\dag)} 
\prod_{x,\mu} \int dU\, e^{\frac{\beta}{N} \re\tr(J_{x\mu}^\dag U)}
\label{eq::Z::1link}
\ee
where $J_{x\mu}$ is a function of the auxiliary variables attached to the link $(x,\hat\mu)$,
\be
J_{x\mu} = \sum_{\nu\neq\mu} \lpar R_{x-\hat\nu,\nu\mu}^\dag Q_{x-\hat\nu,\nu\mu} + R_{x\mu\nu} \rpar
\ee
We refer to \Eq{eq::Z::1link} as the 1-link partition function, given that it only contains $\frac{1}{4}$-plaquette terms consisting of a single link variable coupled to its ambient auxiliary fields. 

The integrand of \Eq{eq::Z::1link} factorizes into a product of one-link integrals, which can be solved analytically \cite{one-link-integral}. For example, the 0-link partition function of compact $U(1)$ lattice gauge theory is given by:
\be
Z \propto \int [dRdR^\conj][dQdQ^\conj]\, 
e^{-\frac{\beta}{2}RR^\conj} 
e^{-\frac{3\beta}{2}QQ^\conj} 
\prod_{x,\mu} I_0(\beta|J_{x\mu}|)
\label{eq::Z::0link::U1}
\ee
where $I_\nu(z)$ are modified Bessel functions of the first kind. We refer to \Eq{eq::Z::0link::U1} as a 0-link partition function, given that all link variables have been integrated out.

For the various $n$-link representations with $n>0$, loop observables can be computed ordinarily as path-ordered products of link variables around the loop $\ell$:  $U(\ell) = \prod_{l\in\ell} U_l$. In the 0-link representation, however, the link variables are also integrated out in the definition of the loop observables. It is straightforward to show that the 0-link loop observables are path-ordered products of effective links $\widetilde U_{x\mu}$: $\widetilde U(\ell) = \prod_{l\in\ell} \widetilde U_l$, so that
\be
\left\langle \frac{1}{N} \tr(U(\ell)) \right\rangle = 
\left\langle \frac{1}{N} \tr(\widetilde U(\ell)) \right\rangle
\ee
The effective links $\widetilde U_{x\mu}$ are defined as the first moments of the corresponding one-link integrals:
\be
\widetilde U^{ij}_{x\mu} = \langle U^{ij} \rangle_{J_{x\mu}}
= \frac{1}{Z} \int dU\, U^{ij}\, e^{\frac{\beta}{N} \re\tr(J_{x\mu}^\dag U)}
\ee
In compact $U(1)$ lattice gauge theory, for example, the effective link is simply given by:
\be
\widetilde U_{x\mu} = \frac{I_1(\beta|J_{x\mu}|)}{I_0(\beta|J_{x\mu}|)}
\frac{J_{x\mu}}{|J_{x\mu}|}
\ee

In Monte Carlo simulations of the $n$-link partition functions, for $n>0$, the $U,Q,R$ variables are updated as follows:
\begin{enumerate}
\item Update $Q_{x\mu\nu}$ via Gaussian heatbath with variance $\frac{N}{\beta}$ and mean $U_{x\mu}U_{x+\hat\mu,\nu} + U_{x\nu}U_{x+\hat\nu,\mu}$;
\item Update $R_{x\mu\nu}$ via Gaussian heatbath with variance $\frac{N}{\beta}$ and mean $Q_{x\mu\nu}U_{x+\hat\mu,\nu}^\dag + U_{x\mu}$;
\item Update $U_{x\mu}$ via (pseudo)heatbath with respect to the Boltzmann weight $e^{\frac{\beta}{N}\re\tr(J_{x\mu}^\dag U_{x\mu}^{})}$
\end{enumerate}
The same algorithm can be used to simulate the 0-link partition function, in which case the link variables $U$ play the role of auxiliary variables, while the $Q,R$ play the role of dynamic variables; and so the $U$ variables must then be updated prior to the $Q,R$ variables.

Numerical simulations confirm that the expectation value of loop operators (e.g. the plaquette operator) is reproduced exactly in all $n$-link representations, in the entire range of $\beta$. However, the more auxiliary fields are present, the larger are the errors associated with lattice measurements, and the larger are the corresponding autocorrelations. In short, the $n$-link representations with $n < 4$ are not adequate for simulations of the pure lattice gauge theory.

\section{Compact lattice QED}

Given that the staggered fermion action is linear in the link variables, the 1-link representation \Eq{eq::Z::1link} of the pure gauge theory can be used to extend the MDP representation of the strong coupling limit \Eq{eq::Z::MDP} to arbitrary values of $\beta$, by shifting $J_{x\mu}$ before performing the group integration:
\be
\frac{\beta}{2N} J_{x\nu} &\mapsto& 
\frac{\beta}{2N} J_{x\nu} + \eta_{x\nu} e^{+a \mu \delta_{\nu\tau}} 
\q_x \qbar_{x+\hat\nu}
\\
\frac{\beta}{2N} J_{x\nu}^\dag &\mapsto& 
\frac{\beta}{2N} J_{x\nu}^\dag - \eta_{x\nu} e^{-a \mu \delta_{\nu\tau}} 
\q_{x+\hat\nu} \qbar_{x}
\ee
Here $\mu$ denotes the quark chemical potential.

\begin{figure}
\centering
\includegraphics[width=0.30\textwidth]{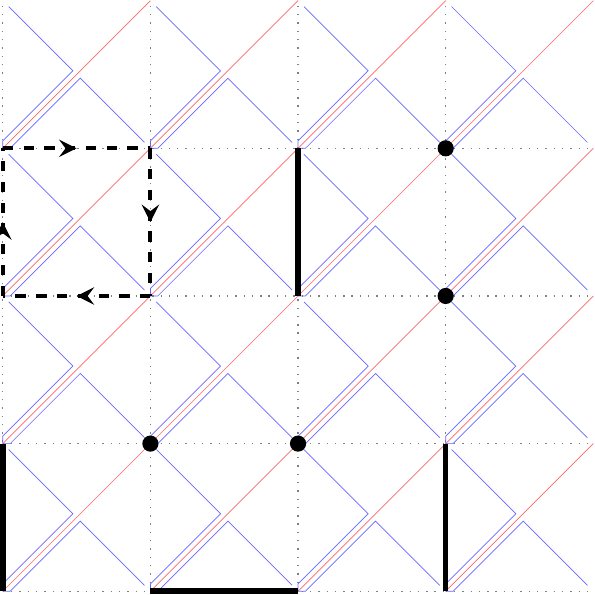}
\caption{Typical MDP configuration of $N_f=1$ compact lattice QED on a $4 \times 4$ lattice,  at finite $\beta$: the red ``diagonal'' links and the blue ``folded'' links represent the auxiliary $Q$ and $R$ variables, respectively; dots represent monomers, solid bonds represent dimers, and dashed lines represent electron loops. \label{fig::mdp_qed}}
\end{figure}

We illustrate this extension with $N_f=1$ compact lattice QED on a lattice torus \cite{future}, for which the electron chemical potential vanishes due to Gauss's law. Its 0-link partition function at finite $\beta$ is given by:
\be
Z &\propto& 
\int [d\qbar d\q]\, e^{2am \sum_{x} \qbar_x\q_x} 
\int [dRdR^\conj][dQdQ^\conj]\, 
e^{-\frac{\beta}{2} RR^\conj} 
e^{-\frac{3\beta}{2} QQ^\conj} 
\prod_{x,\mu} \int dU\, e^{\re((\beta J_{x\mu}^\conj - \eta_{x\mu}\q_{x+\hat\mu}\qbar_x) 
U)}
\nonumber\\
&\propto& 
\int [dRdR^\conj][dQdQ^\conj]\, 
e^{-\frac{\beta}{2} RR^\conj} 
e^{-\frac{3\beta}{2} QQ^\conj} 
\prod_{x,\mu} I_0(\beta|J_{x\mu}|)
\int [d\qbar d\q]\, e^{2am \sum_{x} \qbar_x\q_x} ~\times
\nonumber\\
&&\times~ 
	(1 +  M_x M_{x+\hat\mu} 
	+ \eta_{x\mu} {\widetilde U}_{x\mu}^\conj \q_x\qbar_{x+\hat\mu}
	- \eta_{x\mu} \widetilde U_{x\mu} \q_{x+\hat\mu}\qbar_x)
	\label{eq::Z::0link::QED}
\ee
In addition to the ``meson'' hoppings also present in the strong coupling limit, \Eq{eq::Z::0link::QED} contains $U(1)$-invariant electron hoppings of the form ${\widetilde U}_{x\mu}^\conj \q_x\qbar_{x+\hat\mu}$, which vanish at $\beta=0$. After Grassmann integration, we obtain the MDP representation of compact lattice QED at finite $\beta$:
\be
Z &\propto&
\int [dRdR^\conj][dQdQ^\conj]\, 
e^{-\frac{\beta}{2} RR^\conj} 
e^{-\frac{3\beta}{2} QQ^\conj} 
\prod_{x,\mu} I_0(\beta|J_{x\mu}|)
	\sum_{\{n,k,\ell\}} 
	\prod_x (2am)^{n_x}\,
	\sigma(\ell)\,
	{\widetilde{U}^\conj}(\ell)
	\label{eq::Z::QED::MDP}
\\
&\propto&
	\sum_{\{n,k,\ell\}} 
	\prod_x (2am)^{n_x}\,
	\sigma(\ell)\,
	\langle \widetilde{U}^\conj(\ell) \rangle
	\label{eq::Z::QED::loop}
\ee
In the strong coupling limit, \Eq{eq::Z::QED::loop} reduces to $Z = \sum_{\{n,k\}} \prod_x (2am)^{n_x}$, which sums over monomer-dimer coverings of the spacetime lattice. At finite $\beta$, in addition to monomers and dimers, there are also contributions from self-avoiding electron loops $\ell$, whose weight is given by the pure-gauge expectation value of the Wilson loop around $\ell$, multiplied by an overall sign (see \Fig{fig::mdp_qed}). Such a representation corresponds to an exact loop expansion of the determinant of the Dirac operator. 

This representation has a sign problem due to the sign $\sigma(\ell)$ of the electron loops and the sign of $\re(\widetilde U(\ell))$. In the strong coupling limit, the sign problem vanishes and is therefore expected to be mild at small values of $\beta$, where the typical configuration is a dilute gas of small electron loops in a monomer-dimer bath. For increasing $\beta$, however, the entropy of electron loops dominates, and consequently the system is expected to enter a regime in which the sign problem is severe. Nonetheless, the sign problem is expected to be manageable in some window of small $\beta$.

\section{Summary}

We showed that it is possible to fully decouple the link variables around a plaquette in the Wilson gauge action, at the cost of coupling them to a new set of free auxiliary fields. Consequently, the gauge fields can be integrated out exactly for any value of $\beta$, thus resulting in a representation of the lattice partition function without link variables. In the 0-link representation, however, errors are larger and autocorrelation times are longer than in the original 4-link representation, which renders it unattractive for simulations of the pure gauge theory. 

Nonetheless, it allows us to integrate out the link variables exactly also in the presence of an arbitrary number of (staggered) fermion flavors. We illustrated this with the simple example of compact lattice QED on a torus, for which we obtained an exact MDP representation of its partition function at arbitrary $\beta$, as a sum over electron loop configurations. Numerical simulations using worm algorithms are in progress.

We also plan to construct MDP representations of lattice gauge theories with non-trivial dependence on the chemical potential, e.g. $N_f>1$ lattice QED, and non-Abelian gauge theories including lattice QCD. The corresponding partition functions are obviously more complicated, due to the proliferation of gauge-invariant terms that contribute to their 0-link partition functions. But the recent successful simulations of the strong coupling regime of lattice QCD at finite density in the MDP representation \cite{forcrand-mdp} suggests that a larger window of lattice couplings might be accessible to Monte Carlo simulations using the methods presented in this talk.

\end{document}